\documentclass[11pt]{amsart}
\usepackage{graphicx}
\usepackage{amssymb}  
\usepackage{amsmath}
\usepackage[normalem]{ulem}
\usepackage{enumitem}
\usepackage{hyperref}
\usepackage{a4wide}
\usepackage{bm}
\usepackage{amsthm}

\theoremstyle{definition}

\newcommand{\Dn}{\mathcal{D}_n}

\newcommand{\cD}{\mathcal{D}}
\newcommand{\cA}{\mathcal{A}}

\newcommand{\cM}{\mathcal{M}}

\newcommand{\cS}{\mathcal{S}}
\newcommand{\Mat}{\mathrm{Mat}}

\newcommand{\tr}{\mathrm{tr}}

\newcommand\CC{\mathbb{C}}

\newcommand\ess{\mathbf{s}}

\newcommand\zee{\mathbf{z}}

\begin{document}

\title{A symmetry-inclusive algebraic approach to genome rearrangement}

\author{Venta Terauds, Joshua Stevenson and Jeremy Sumner}

\address{Discipline of Mathematics, School of Natural Sciences, Private Bag 37: University of Tasmania, Sandy Bay, Tasmania 7001, Australia}

\email{venta.terauds@utas.edu.au\\joshua.stevenson@utas.edu.au\\ jeremy.sumner@utas.edu.au  }

\thanks{This work was supported by Australian Research Council Discovery Grant DP180102215, by an Australian Government Research Training Program (RTP) Scholarship (second author) and by use of the Nectar Research Cloud, a collaborative Australian research platform supported by the NCRIS-funded Australian Research Data Commons (ARDC).}

\begin{abstract}
Of the many modern approaches to calculating evolutionary distance via models of genome rearrangement, most are tied to a particular set of genomic modelling assumptions and to a restricted class of allowed rearrangements. The ``position paradigm'', in which genomes are represented as permutations signifying the position (and orientation) of each region, enables a refined model-based approach, where one can select  biologically plausible rearrangements and assign to them relative probabilities/costs. Here, one must further incorporate any underlying structural symmetry of the genomes into the calculations and ensure that this symmetry is reflected in the model. 
In our recently-introduced framework of {\em genome algebras}, each genome corresponds to an element that simultaneously incorporates all of its inherent physical symmetries. The representation theory of these algebras then provides a natural model of evolution via rearrangement as a Markov chain. Whilst the implementation of this framework to calculate distances for genomes with `practical' numbers of regions is currently computationally infeasible, we consider it to be a significant theoretical advance: one can incorporate different genomic modelling assumptions, calculate various genomic distances, and compare the results under different rearrangement models. The aim of this paper is to demonstrate some of these features. 
\end{abstract}

\keywords{genome rearrangement models; genome symmetry; evolutionary distance; representation theory}

\maketitle

\section{Introduction}

Genome rearrangement modelling has historically been approached as a combinatorial problem, with the aim of developing fast algorithms to compute pairwise distances between genomes. Although permutations have long been used to represent genomes, it is only this century that serious consideration has been given to the algebraic frameworks that form the theoretical basis for the models. Beginning with the work of Meidanis and Dias \cite{mei-dias}, and continuing with many others \cite{feijao-mei-12,andrew14,attilaand,bhatia-15}, it has been recognised that developing the algebraic formalism is key to making progress in genome rearrangement modelling, in particular in enabling more refined model-based approaches. 

As classified by Bhatia et al \cite{wildcrazy} in their excellent overview, algebraic frameworks for modelling genome rearrangement tend to use either the ``content'' or the ``position'' paradigm. In the former paradigm, genomes are represented in terms of the adjacencies between regions; in the latter, positions as well as regions are labelled, and genomes are denoted by maps that link regions to positions. Whilst the latter necessarily applies a choice of reference frame in the position labelling (unless the genome possesses no symmetry), the content paradigm has the advantage of being `orientation free', since it only `notices' which regions are adjacent, and in which orientation. 

To model rearrangements in the content paradigm, the ``double cut and join'' \cite{bhatia-15} and, more generally, ``$k$-break'' \cite{feijao-mei-12} operations are widely used. These operations cover inversions, fissions, fusions and translocations, and the ability of a single operation to model a range of events, over multiple chromosomes, has been considered an advantage \cite{bergeron-etal-06}. The double cut and join framework continues to be adapted in various ways, for example to include insertions and deletions \cite{braga-etal-11,bohnen-etal-21}, incorporate intergenic regions \cite{simonaitis-etal,fertin-etal-17}, and limit rearrangements to very small scale events \cite{oliveira-etal-18}. However, these approaches almost always utilise minimum distance as the distance measure (or its generalisation to the median distance \cite{zanetti-median-2016}), and the classification of rearrangements in these approaches remains coarse.

Biologically, the relative probabilities of different rearrangements are likely to differ according to their type (for example, inversions or translocations), size, and position on the genome \cite{darling08,alexeev15}. Utilising the position paradigm framework enables a fine-grained approach to rearrangement models that can incorporate such information: rearrangements can be represented as permutations of positions, that is, operators that switch around the regions that are in particular positions, whatever the regions may be.

The implicit choice of reference frame in this paradigm means that (usually) more than one permutation will represent the same genome, due to inherent symmetry. The inclusion of genome symmetry in the theoretical framework has been previously considered \cite{attilaand}, but in practice this has been added in as a separate element of calculations, thus increasing the computational complexity, whether treating genomes with unsigned regions or signed \cite{mles,galv-baud-dias-17}. That is, the the process has generally been to (i) perform calculations for the genomes as `fixed orientation entities'' (usually group elements) and then (ii) repeat for each of the symmetries to get the result.
 
Less comprehensively considered has been any corresponding symmetry of rearrangement models. Although one chooses a reference frame in order to represent the genomes and rearrangements of interest, the set of allowed rearrangements should be independent of the reference frame. Rearrangement models used in practice have tended to have this property as a consequence of being quite general, for example models consisting of all inversions, all inversions of size $k$, all inversions with some probability $p$ and all transpositions with probability $q$, and so on. However, until recently, \cite{vjez_circ1,vjez_algebras} this has not been included in the theoretical framework.

The approach we present here incorporates the symmetry of genomes and rearrangements in the unified framework of {\em genome algebras}; here each genome and each rearrangement corresponds to a single mathematical object that simultaneously incorporates all of its inherent physical symmetries. Our framework easily incorporates different genomic modelling assumptions and different sets of allowed rearrangements; further, it naturally facilitates the calculation of different measures of evolutionary distance. 

The aim of this paper is to demonstrate some of these features. We outline the construction of the relevant genome algebras for signed circular single-chromosome genomes with and without an origin of replication, and then give some results for computations of genomic distance --- as estimated via minimum distance, mean first passage time, and the maximum likelihood estimate of time elapsed --- under various rearrangement models.

\section{Genome instances and permutation clouds}

The genome algebra framework was presented in Terauds and Sumner \cite{vjez_algebras}, along with details of the construction for the case of unsigned circular genomes with dihedral symmetry. Here we consider the algebra for circular single-chromosome genomes with oriented regions, both with and without an origin of replication. We note that the former construction may also be applied to model linear genomes.

Modelling genomes with oriented regions as elements of the hyperoctahedral group is standard; we refer to Bhatia et al \cite{wildcrazy} and Egri-Nagy et al \cite{attilaand} for detailed treatments. 
We consider that the genomes of interest share $n$ regions in common, where each region is a contiguous section of DNA (these may also be referred to as synteny blocks or conserved regions). Labelling the positions and regions both by $1,\ldots , n$, we represent an {\em instance} of a genome by a signed permutation $\sigma$, mapping regions to positions, where
\[  \sigma(i)= \pm j \; \iff \; \textrm{ region } i \textrm{  is in position } j \textrm{ with positive/negative orientation}\,.\]
Setting $\sigma(-i) = -\sigma(i)$ for all $i$ makes $\sigma$ an element of the hyperoctahedral group $H_n$, modelled as a subgroup of the symmetric group on $\{\pm 1,\ldots , \pm n\}$.  We shall henceforth use the convention of writing $\overline{i}$ instead of $-i$.

With the term {\em instance}, we are emphasising that a single permutation represents an observation of the genome with a fixed physical orientation and a choice of position labelling. The labelling of the regions (including region orientation) is chosen once and is immutable, however, the labelling of the positions reflects a choice. For a genome with an origin of replication, we may decide on a labelling such that the origin lies between positions $1$ and $n$. However, there remains a choice: we may label the positions either clockwise or anticlockwise. Thus there are two distinct permutations that may represent any given genome. For a fixed reference frame, this corresponds to the two physical orientations of the genome obtained by flipping it over in space. Thus, in this case, the symmetry group corresponding to the genomes has size two (it's $\cS_2 = \{e,f\}$ -- a ``do nothing'' and a ``flip''), and we equivalently say that each genome has two instances, represented by the group elements $\{\sigma, f\sigma\} = \cS_2\sigma$. 
A circular genome with no distinguished position has dihedral symmetry (one can rotate as well as flip);  the symmetry group is a copy of the dihedral group, $\Dn$ and each genome thus has $2n$ distinct instances, corresponding to elements of a coset $\Dn\sigma$.

Similarly, we model an instance of a rearrangement as a signed permutation that acts on a genome instance on the left, mapping signed positions to signed positions. For example, an inversion of the regions in positions 1 and 2 would be $a = [\overline{2},\overline{1},3,4,\ldots,n]$ (expressed in one-line notation). As with genomes, there are $|Z|$ instances of any given rearrangement, where $Z$ is the relevant symmetry group.\footnote{Note that, as rearrangements, these need not all {\em act} distinctly.} 
In earlier work \cite{vjez_circ1},we considered model symmetry as a two step process: if $a$ above were an allowed rearrangement of genomes with origin of replication between positions $n$ and $1$, then the rearrangement that swaps the regions in positions $n$ and $n-1$, namely $[1,2,3,\ldots,\overline{n},\overline{n-1}]$, should be allowed and assigned the same probability.  

In the {\em genome algebra}, all of the instances for a given genome (or rearrangement) are combined into a single element: a genome corresponds to the sum of its instances, each weighted by $1/(\#\textrm{symmetries})$. Such elements are the {\em basis elements} of the genome algebra, meaning that everything in the genome algebra may be expressed as a linear combination of genomes. In the genome algebra, elements may be added, multiplied by scalars and multiplied together; in particular, the latter is how we model a rearrangement acting on a genome. 

To formally construct the genome algebra we begin with a group $G$, whose elements represent instances of the genomes, and a subgroup $Z\subseteq G$, representing their symmetries, and form the {\em symmetry element} 
\[ \zee := \tfrac{1}{|Z|} \sum_{z\in Z} z \,\]
of the group algebra $\CC[G]$.\footnote{Recall that the group algebra is formed from all finite linear combinations of group elements.} The genome algebra of $G$ with $Z$ is $\cA:= \zee\CC[G]$. The distinct genomes with instances in $G$ correspond to the distinct elements of the set 
$\{ \zee \sigma : \sigma \in G\}$, 
which forms a basis for $\cA$.  
For example, taking $G=H_n$ and $Z=\cS_2$, the symmetry element is $\zee = \frac{1}{2} e + \frac{1}{2} f$ and the distinct genomes have the form $\zee\sigma =\frac{1}{2}\sigma + \frac{1}{2}f\sigma$ for $\sigma\in H_n$. One can think of a genome as existing as the average of its instances, where there is an equal probability of observing any particular instance. 

Now rearrangements also have the form $\zee a$, for $a\in G$, and rearrangement occurs via left action on a genome, $\zee\sigma\mapsto\zee a\cdot \zee\sigma$, which we can think of as ``all orientations of ($a$ acting on (all  orientations of $\sigma$))''. This results in a linear (convex) combination of genomes; in fact, this is a probability distribution of the genomes that may result.

As an explicit example, consider the following reference genome (with instance $e$) in the genome algebra $\cA$ for the group $H_6$ with symmetry group $\cS_2$:
\[ \zee e = \zee = \tfrac{1}{2}\left( \raisebox{-9mm}{\includegraphics[scale=0.4]{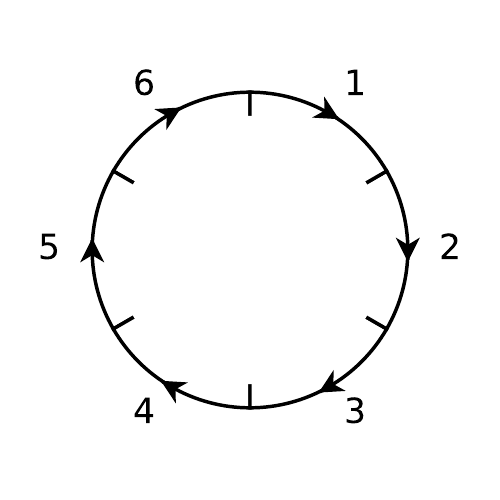}}+  \raisebox{-9mm}{\includegraphics[scale=0.4]{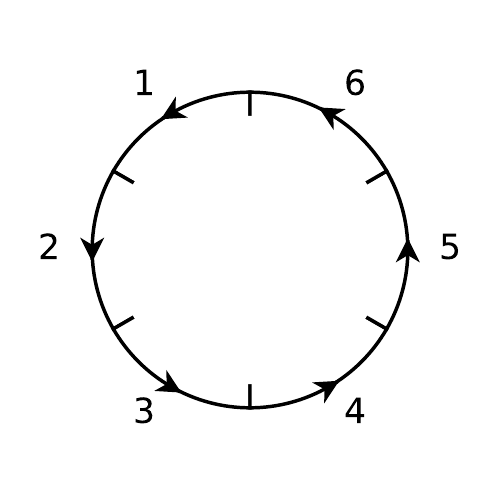}}\right)\,. \]
Choosing the rearangement instance $a = [\overline{2},\overline{1},3,4,\ldots,6]$ and applying the rearrangement $\zee a$ to the reference genome $\zee$ in the genome algebra, we obtain
\begin{eqnarray*}
(\zee a)\cdot \zee &=& \tfrac{1}{4}\left(  \raisebox{-9mm}{\includegraphics[scale=0.4]{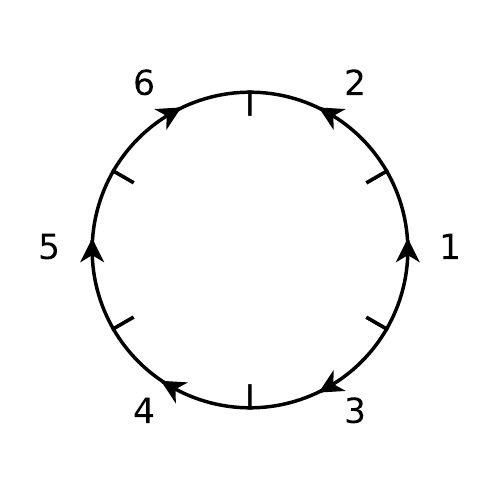}} + \raisebox{-9mm}{\includegraphics[scale=0.4]{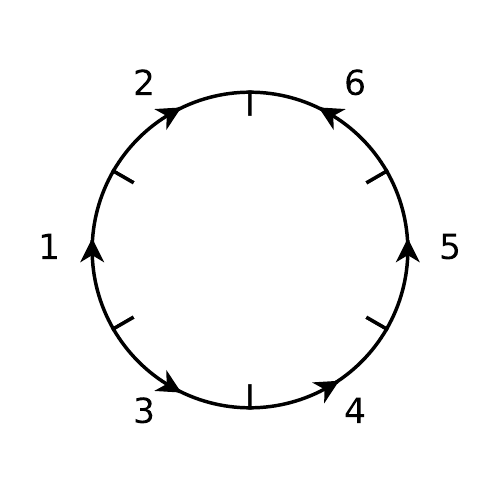}}
		+   \raisebox{-9mm}{\includegraphics[scale=0.4]{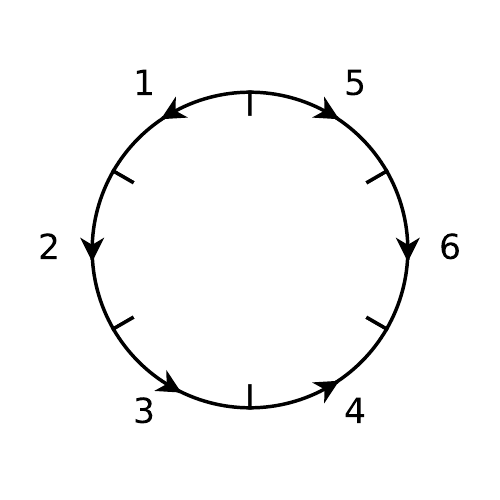}}+   \raisebox{-9mm}{\includegraphics[scale=0.4]{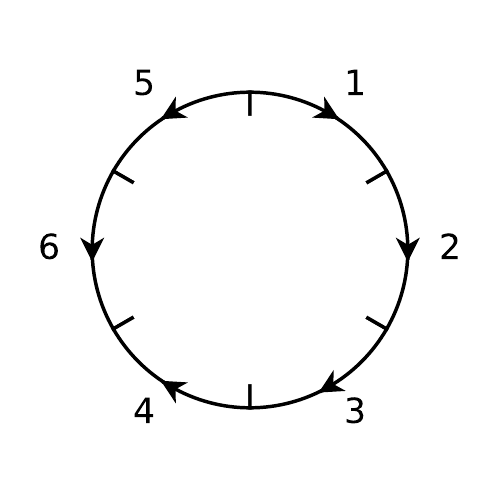}}\right)\\
		&=& \tfrac{1}{2} \left( \zee \cdot \raisebox{-9mm}{\includegraphics[scale=0.4]{ae.pdf}} + \zee \cdot  \raisebox{-9mm}{\includegraphics[scale=0.4]{af.pdf}}\right)
		= \tfrac{1}{2}\zee a +\tfrac{1}{2}\zee \sigma\,,
\end{eqnarray*}
where $\sigma = af$. This means that, by applying the rearrangement $\zee a$ to the reference genome $\zee$, one would obtain either the genome $\zee a$ or the genome $\zee\sigma$, with equal probabllities. 

In order to compute evolutionary distance via rearrangement, we begin by fixing a model. Formally, a model  is a set 
\[ \cM:= \{ \zee a_1\zee, \ldots , \zee a_q\zee \} \subseteq \cA\]
for some $a_1,\ldots a_q\in G$, along with a probability distribution $w:\cM\to (0,1]$. We note that, whilst there are exactly $\tfrac{|G|}{|Z|}$ distinct genomes $\zee\sigma$ that have instances in $G$ and symmetry group $Z$ (for example, this is $2^{n-1}n!$ for signed genomes with an origin of replication and $2^{n-1}(n-1)!$ for those with no distinguished position), there are fewer distinct rearrangements. In particular, since $\zee$ is an idempotent ($\zee\cdot\zee = \zee$), rearrangments $\zee a $ and $\zee b$ have the same left action on genomes whenever $\zee a\zee = \zee b\zee$. This motivates the above formulation of the model, since unintentionally duplicating a particular rearrangement action could unintentionally skew the probability distribution. In Section~\ref{sec:results}, we expand on this and provide an example.

Given a model $(\cM,w)$, we form the {\em model element} in the genome algebra:
\[ \widetilde{\ess}: = \sum_{\zee a \zee \in \cM} w(\zee a \zee) \zee a\,.\]
Since $w$ is a probability distribution, $\widetilde{\ess}$ is a convex sum, and thus, in a direct extension of the above, we see that acting on a genome on the left with the model element, $\zee\sigma\mapsto\widetilde{\ess}\cdot\zee\sigma$, results in a convex combination of genomes. These are exactly the genomes that may be obtained from $\zee\sigma$ via one rearrangement chosen from the model, with their respective probabilities given by the coefficients. The {\em left regular representation} of the model element in $\cA$, $\rho(\widetilde{\ess})$, summarises this information in the form of a $K\times K$ matrix (where $K:=\tfrac{|G|}{|Z|}$): labelling the distinct genomes $\{ \zee\sigma_1,\ldots ,\zee\sigma_K\}$ ,
\[ \rho(\widetilde{\ess})_{ij} := \textrm{coefficient of } \zee\sigma_i \textrm{ in expansion of } \widetilde{\ess}\cdot\zee\sigma_j \,,\]
and thus
\begin{eqnarray*} \left(\rho(\widetilde{\ess})^k\right)_{ij} =  \rho(\widetilde{\ess}^k)_{ij}
	&=& \textrm{probability of } \zee\sigma_j \mapsto \zee\sigma_i \textrm{ via } k \textrm{ rearrangement events}\,.
\end{eqnarray*}
We thus see evolution via rearrangement explicitly as a discrete Markov process, with transition matrix $\rho(\widetilde{\ess})$.
We define the {\em path probabilities} for a genome $\zee\sigma_i$ with respect to the reference genome $\zee$ via
\[ \alpha_k(\zee\sigma_i) := \textrm{ probability of obtaining } \zee\sigma_i \textrm{ from } \zee \textrm{ via } k \textrm{ rearrangement events} \,.\]
Now, these can be read from the first column of the matrix $\rho(\widetilde{\ess})^k$, or obtained via the trace of a modified matrix \cite{vjez_algebras}:
\begin{equation}\label{eq:alphas-trace}
\alpha_k(\zee\sigma) = \tfrac{|Z|}{|G|}\tr(\rho(\widetilde{\ess}^k\zee\sigma^{-1})) = \tfrac{1}{K}\chi(\widetilde{\ess}^k\zee\sigma^{-1})\,.
\end{equation}
To compute this more efficiently, one would usually utilise the algebra's  irreducible modules and representations. 
As with the regular representation, each irreducible {\em representation} is a linear mapping that assigns to each algebra element a matrix, in such a way that preserves multiplication. A representation $\rho:\cA\to \Mat_m(\CC)$ defines an action of $\cA$ on the {\em module} $\CC^m$; the module and the representation are {\em irreducible} if $\CC^m$ contains no non-trivial subspace invariant under this action. The irreducible representations always exist and enable the computation of the path probabilities via matrices of much smaller dimension than the regular representation\footnote{although in this case the algebra is not isomorphic to to a direct sum of its irreducible modules; see \cite{vjez_algebras} for full details}.

We note that we would usually assume that the model is sufficient to generate the set of genomes, that is, that the Markov chain is irreducible. When one considers genomes and rearrangements simply as group elements, the analogue of this condition is that the permutations in the model generate the entire group \cite{attilaand,mles}. In fact, these formulations are equivalent \cite{josh_us}.
The only other condition that we put on the model is {\em model reversibility}, that is, that whenever $\zee a \zee\in \cM$, one must have $\zee a^{-1}\zee \in\cM$ with $w(\zee a\zee) = w(\zee a^{-1}\zee)$. This is not required for any of the above, however it ensures that the Markov process is reversible \cite{vjez_algebras}, so that $\sigma^{-1}$ can be replaced by $\sigma$ in (\ref{eq:alphas-trace}). Further, it allows the path probabilities to be more efficiently computed via diagonalising the respective matrices. 

The path probabilities, or more generally, the Markov matrix, can be utilised to compute various types of evolutionary distance measures. We consider three of these in the next section.

\section{Evolutionary distance}

Genome evolution via rearrangement is most commonly modelled as a discrete process -- that, is a sequence of discrete rearrangement events -- and the most ubiquitous measure of evolutionary distance is the minimum distance. This usually takes the form of the minimum number of rearrangements that can be applied to the reference genome to obtain the target, but also includes minimum weighted distance, where different costs are applied to different types of rearrangement \cite{baudet15,nguyen-et-al-2005}. The issues with minimum distance as a proxy for true evolutionary distance have been widely discussed in the literature and many alternative measures have been proposed; these are often based either on modifying minimum distance or breakpoint distance in some way \cite{Wang_Warnow_06,Lin_Moret}. 

The maximum likelihood estimate of time elapsed (MLE), introduced by Serdoz et al \cite{mles}, takes a completely different approach to evolutionary distance. Here, evolution is modelled as a discrete sequence of rearrangement events occuring in continuous time. The Poisson distribution is a natural choice for the distribution of events in time, since combining it with the discrete time Markov chain above produces the corresponding continuous time Markov chain,
\[ P(t):= e^{\left(\rho(\widetilde{s})-I\right)t}\,\]
which encapsulates the likelihoods. To be specific, for a genome $\zee\sigma$, the likelihood function at a time value $t$ is the probability that the reference genome $\zee$ evolved into $\zee\sigma$ in time $t$, that is,
\[ L(\zee\sigma|t)= P( \zee\mapsto\zee\sigma \textrm{ in time } t) =\sum_{k\geq 0} \alpha_k(\zee\sigma) P(k \textrm{ events in time } t)\,. \]
The MLE distance from $\zee$ to $\zee\sigma$ is then the value of $t$ for which the likelihood function attains a maximum. 

As has been noted \cite{vjez_circ1,vjez_algebras}, one could choose a different distribution to combine with the path probabilities and calculate likelihood functions.  
However, utilising the above, the likelihood function for a given genome $\zee\sigma$ can be found via the matrix trace \cite{jezandpet,vjez_algebras}. 
We note that the MLE does not always exist. This may be seen as a feature \cite{mles,vjez_circ1} --- it is biologically realistic that not all pairs of genomes display an evolutionary signal, and hence a finite evolutionary distance, under a given model --- or a fault; see Francis and Wynn \cite{francis-wynn-20} for more discussion of this. 

One of the motivators for considering the MLE is that it takes into account all possible paths via rearrangement between two genomes, along with their probabilities. For example, there may be a minimum path of length $k$ rearrangements between two given genomes,   but many more possible paths of length $\ell$ between them, making evolution via $\ell$ rearrangements the more {\em likely} evolutionary scenario. The discrete Markov model and accompanying depiction of the genome space as an edge-weighted graph (an algebra version of a Cayley graph) allow other distance measures that incorporate this information to be considered. Such graphs have previously been utilised for genomes modelled as group elements \cite{moulton-steel-12,attilaand,clark-etal-19,francis-wynn-20} and our framework extends this to algebra elements, where symmetry is automatically included. 

One such measure is {\em mean first passage time} (MFPT), a well-studied concept in Markov chain theory whose application to genome rearrangement was recently considered by Francis and Wynn \cite{francis-wynn-20}. The mean first passage time is the expected length of a random walk on the edge-weighted graph that starts at the reference and ends when it first encounters the target genome. Extending Francis and Wynn's treatment from the group case to the genome algebra framework, the mean first passage time distance may be calculated directly from the  Markov matrix $\rho(\widetilde{\ess})$ via a simple row replacement and matrix inversion \cite{francis-wynn-20}.

\section{Some example computations}\label{sec:results}

To demonstrate the flexibility one has in applying the genome algebra framework, we present some results from computations for genomes with six oriented regions. 
We claim no algorithmic sophistication, and have not applied any numerical methods to speed up computation. Thus, due to the size of the matrices involved, the results available at the time of submission are primarily for genomes modelled with no distinguished position (there are $23,040 = 2^5 6!$ distinct genomes with a distinguished position on 6 regions, and $3840 = 2^5 5!$ without). 
Our small sample of results shows that changing either the rearrangement model or the distance measure can greatly affect the relative genomic distances obtained. Thus, whilst computations for large numbers of regions remain currently out of reach, the framework has the potential to provide insights in this vein.

With the introduction of the MLE distance measure in Serdoz et al \cite{mles}, examples were provided of the MLE and the minimum difference measures ordering genomes with unsigned regions and dihedral symmetry differently, in terms of their distance from the reference. In our subsequent work \cite{vjez_circ1}, we gave further examples of this, along with an example of the MLEs calculated under two different models ordering (unsigned) genomes differently. 
Here we investigate the differences betwen the following models:
\begin{enumerate}
\item[(i)] inversions of one and two regions; equally likely
\item[(ii)] inversions of one and two regions; single region twice as likely
\item[(iii)] ``all inversions equally likely''
\item[(iv)] inversions of one, two and three regions; equally likely
\item[(v)] inversions of one and two regions and one region translocations; equally likely.
\end{enumerate}

By ``all inversions'' in model (iii), we mean all inversions of regions of up to size 5 (since an inversion of all six regions is just a flip). This may naively seem correct; however this is in fact a duplication of rearrangements, and we include it here in this way to demonstrate this effect. Note that any instance of an inversion of four regions, written as $a\in H_6$, is the `flip' of an inversion of $6-4=2$ regions, that is, $a=d b$  (where $d\in D_6$) for an instance $b\in H_6$ of an inversion, and thus we include $\zee a = \zee b$ twice in the model. The same applies to inversions of sizes $1$ and $5$. We can think of this as the inversions $a$ and $b$ being {\em complementary}; in any case, including these duplicate rearrangements skews the probablity distribution, and rather than obtaining the intended uniform distribution, the result is a model that has inversions of sizes $1$ and $2$ with twice the probability of inversions of size $3$. Thus, for 6 regions, model (iv) is the correct implementation of an ``all inversions equally likely'' model. 

All of our computations were performed on an instance of the Nectar Research Cloud running Ubuntu 18.04 with 32 gigabytes of available RAM. We found the $3840\times 3840$ Markov matrix $\rho(\widetilde{\ess})$ for each model using SageMath \cite{sage}, and used this to calculate the MFPT distances (matrix computation in SageMath, assuming a uniform mean inter-arrival time) and the minimum distances (via the Cayley graph of the matrix and the\texttt{ nx.shortest\_path\_length} function in the Python package  Networkx \cite{networkx}). For the MLEs, we found the relevant  irreducible representation matrices of $H_6$ via SageMath (utilising the Gap \cite{GAP4} package repsn \cite{repsn}) and projected these onto the irreducibles of the genome algebra in order to compute the likelihood functions via the irreducible characters (see our paper \cite{vjez_algebras} for more details); we then used an optimising function to find the MLE (or that there wasn't one).

Since the genomes form equivalence classes --- in particular, all of these distances are the same for genomes $\zee\sigma$ and $\zee\tau$ whenever $\sigma = \tau^{-1}$ or $\zee\sigma\zee$ = $\zee\tau\zee$ \cite{vjez_algebras} --- we needed only calculate the distances from the reference genome to 250 representatives to have the pairwise distances between all pairs of genomes. (As usual, if the distance between the reference and $\zee\sigma$ is $d(\zee,\zee\sigma)$, then the distance between $\zee\sigma_1$ and $\zee\sigma_2$ is $d(\zee,\zee\sigma_1\sigma_2^{-1})$.) 
We highlight a few observations from these results.

\begin{table}[h]
\centering
\caption{Pairwise distance estimates via each of minimum distance, MLE and MFPT, between genomes with instances $e=[1,2,3,4,5,6], \sigma_1=[3,4,1,\overline{2},6,5],\sigma_2=[\overline{6},1,2,5,4,3]$, assuming dihedral symmetry, under the five rearrangement models.}
$
\begin{array}{lc|c|c|c}
d & \textrm{model(s)} & d(\zee,\zee\sigma_1) & d(\zee,\zee\sigma_2) & d(\zee\sigma_1,\zee\sigma_2) \\
\hline
 & (i),(ii) & 5 & 4 & 5\\
\textrm{min} & (iii),(iv) & 4 & 3 & 3\\
 & (v) & 3 & 4 & 3\\
\hline 
 & (i) & - & - & -\\
  & (ii) & - & - & -\\
\textrm{MLE}   & (iii) & 8.659 & 8.81 & -\\
 & (iv) & 6.94 & 4.254 & 7.01\\
  & (v) & - & - & 8.097\\
\hline
& (i) & 4275.0 & 4274.8 & 4277.6\\
  & (ii) & 4354.4 & 4361.5 & 4361.7\\
 \textrm{MFPT}  & (iii) & 4159.0 & 4159.1 & 4158.4\\
 & (iv) & 4193.8 & 4188.2 & 4193.6\\
  & (v) & 3992.1 & 3994.0 & 3991.0\\
\end{array}
$
\end{table}

Considering the MFPT distances, we observe that adding weights to model (i) to favour the smaller inversions (model (ii)), changes the ordering of genomes $\zee\sigma_1$ and $\zee\sigma_2$ with respect to their distance from the reference genome $\zee = \zee e$. Similarly, there are different relative orderings under the `unintended'  skewed probability distribution of model (iii) and the `true' all inversions equally likely model (iv). Model (v), which includes small translocations, seems to differentiate the MFPT distances the most. We note that the relatively large values of the MFPT distances compared to the MLE and min distances reflect that they are calculated as the mean (weighted) length of a random walk on a quite sparsely connected graph with 3840 nodes. Thus, mean path lengths of the order of the number of nodes are quite reasonable.

For the MLE distance, we observe that there is no detectable evolutionary relationship for any of these pairs of genomes under models (i) and (ii). Under model (iii), the MLE distances of genomes $\zee\sigma_1$ and $\zee\sigma_2$ from the reference are similar (as are the MFPTs), but these two measures order them differently than the minimum distance. 
Under each of the models, we obtained an MLE value for between approximately thirty five and forty per cent of the genomes (although not the same ones in each case). For comparison, previous calculations of MLE distances \cite{mles,vjez_circ1} for genomes with unoriented regions have found a detectable evolutionary signal for around forty five percent of genomes (compared to a fixed reference). We note that the determination of whether or not an MLE exists is highly sensitive to the optimisation function applied.

It is perhaps also interesting that, under model (iv), there is the most `agreement' between the distance measures, in that the MLE, MFPT and minimum distance all order genomes $\zee\sigma_1$ and $\zee\sigma_2$ the same in terms of their distance from the reference. 
 Although the minimum distance clearly has the least resolution (with maximum minimum distances respectively 7, 6 and 5 for each of the three cases), it nonetheless gave different orderings of genomic distances under different models. 

Overall, we found both MLE and MFPT to generally increase with minimum distance, with variation between the models. The MLE displays much more variance than the MFPT overall, which we would assume is due to the inclusion of the stochastic component. We provide plots to illustrate this in \ref{app:6reg}. Further, in \ref{app:3reg} we present constructions and results for all genomes with three regions under different symmetry assumptions.

\section{Conclusion}

The position paradigm approach to genomic modelling enables a fine-grained consideration of rearrangement models, allowing different types of rearrangements, of different scales, and at different genomic positions, to be included in models with different relative probabilities. In this framework, any structural symmetry of the genomes needs to be incorporated into the modelling, which has previously necessitated an extra step in calculations. Here, we have presented a {\em genome algebra} framework that provides a unified approach to the symmetry of genomes and rearrangement models. Our approach reflects biological reality --- objectively, a genome {\em is} a physical object that exists with all of its possible symmetries simultaneously --- and reduces the complexity of the computation process, since the symmetry of the objects is included from the start. 

The sample computations we have provided here are intended to demonstrate the flexibility of our approach in incorporating different models of genomic structure and different rearrangement models, enabling comparisons of the results obtained for different models and different measures of genomic distance. We consider that our symmetry-inclusive approach represents a significant theoretical advance in genome rearrangement modelling. Much work remains, however, to implement the theory in practically useful computations of genomic distance. Whilst the genome algebras represent a dimensionality reduction from previous group-based approaches, the number of distinct genomes is still factorial ($2^{n-1}(n-1)!$ for signed genomes with $n$ regions and dihedral symmetry, for example). This means that the dimension of the  matrices we use for computation is very large, although here, as observed in Francis and Wynn\cite{francis-wynn-20} the Markov matrices are quite sparse, which should make more efficient matrix methods applicable.

Along with pursuing efficient algorithmic and numerical methods, in particular for the calculation of MLEs, we are interested in broadening the application of our framework. In future work, we aim to incorporate insertions and deletions into the framework by extending it to include semigroups, and investigate the potential to model multiple chromosomes and intergenic regions.

\appendix

\section{Further results for the 6-region case}\label{app:6reg}

The following plots are included to give an idea of the range of values taken by the MLE and MFPT distance measures. Since there are (c.f. Section~\ref{sec:results}) 250 equivalence classes of genomes, with the members of each class equidistant from the reference, there are 250 values for the MFPT distance for each model. There are fewer values for the MLE distance, since this does not always exist. We note again that the determination of MLEs is highly dependent on the optimisation function used.

\begin{figure}[ht]
\begin{center}
\caption{Plot of MLE distances ($y$-axis) between each genome and the reference, genomes ordered by min distance and MLE value for model (iii).  Key: model (i) blue; model (ii) red; model (iii) yellow; model (iv) green; model (v) orange.}
\includegraphics[width=0.81\textwidth]{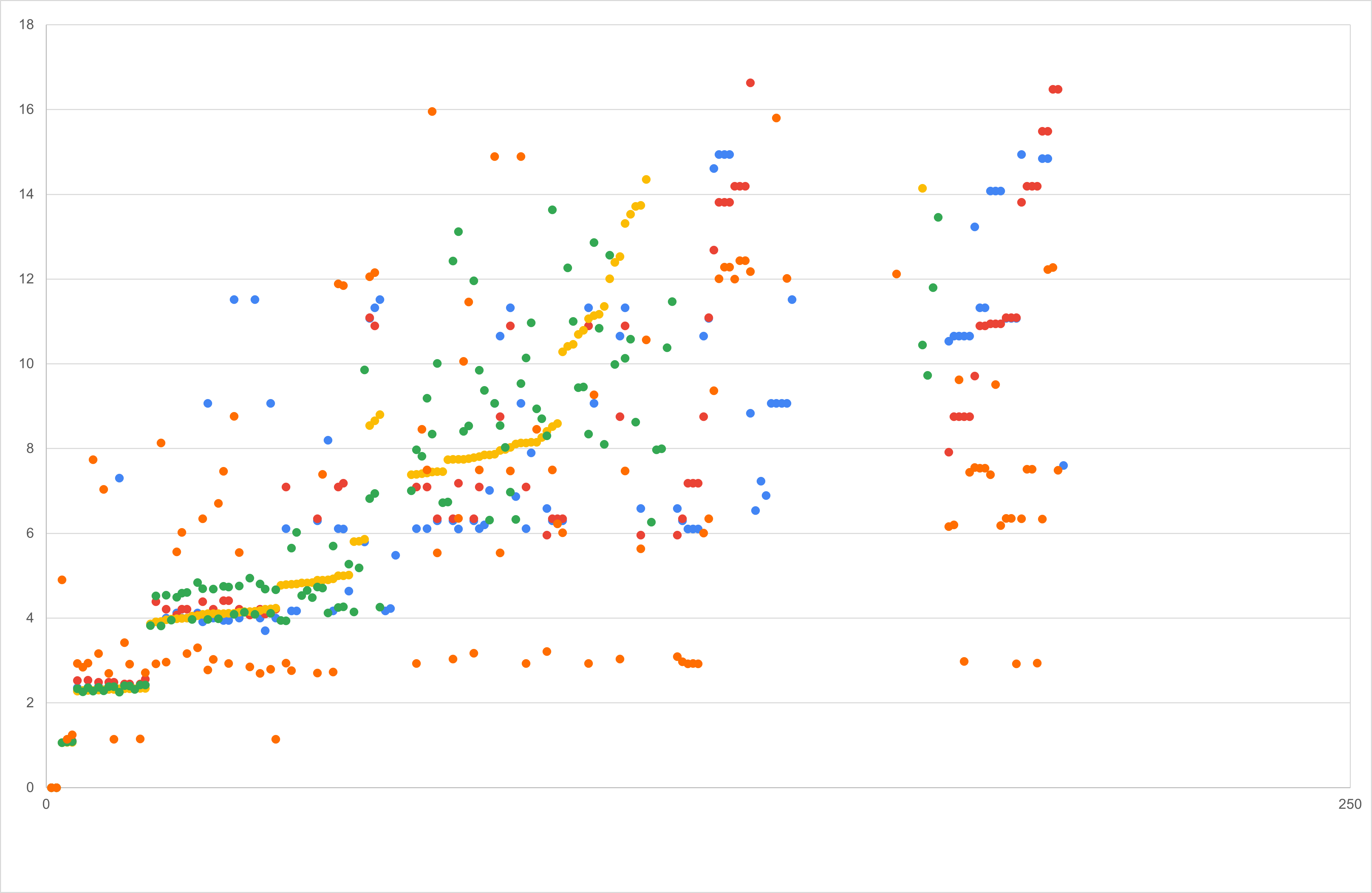}
\end{center}
\end{figure}

The generally lower values taken by the MFPT distance under model (v) reflect the larger number of rearrangements under this model, meaning that the graph is more connected and there are more shorter paths available.

\begin{figure}[ht]
\begin{center}
\caption{Plot of MFPT distances ($y$-axis) between each genome and the reference, genomes ordered by min distance and MLE value for model (iii). Key: model (i) blue; model (ii) red; model (iii) yellow; model (iv) green; model (v) orange.}
\includegraphics[width=0.81\textwidth]{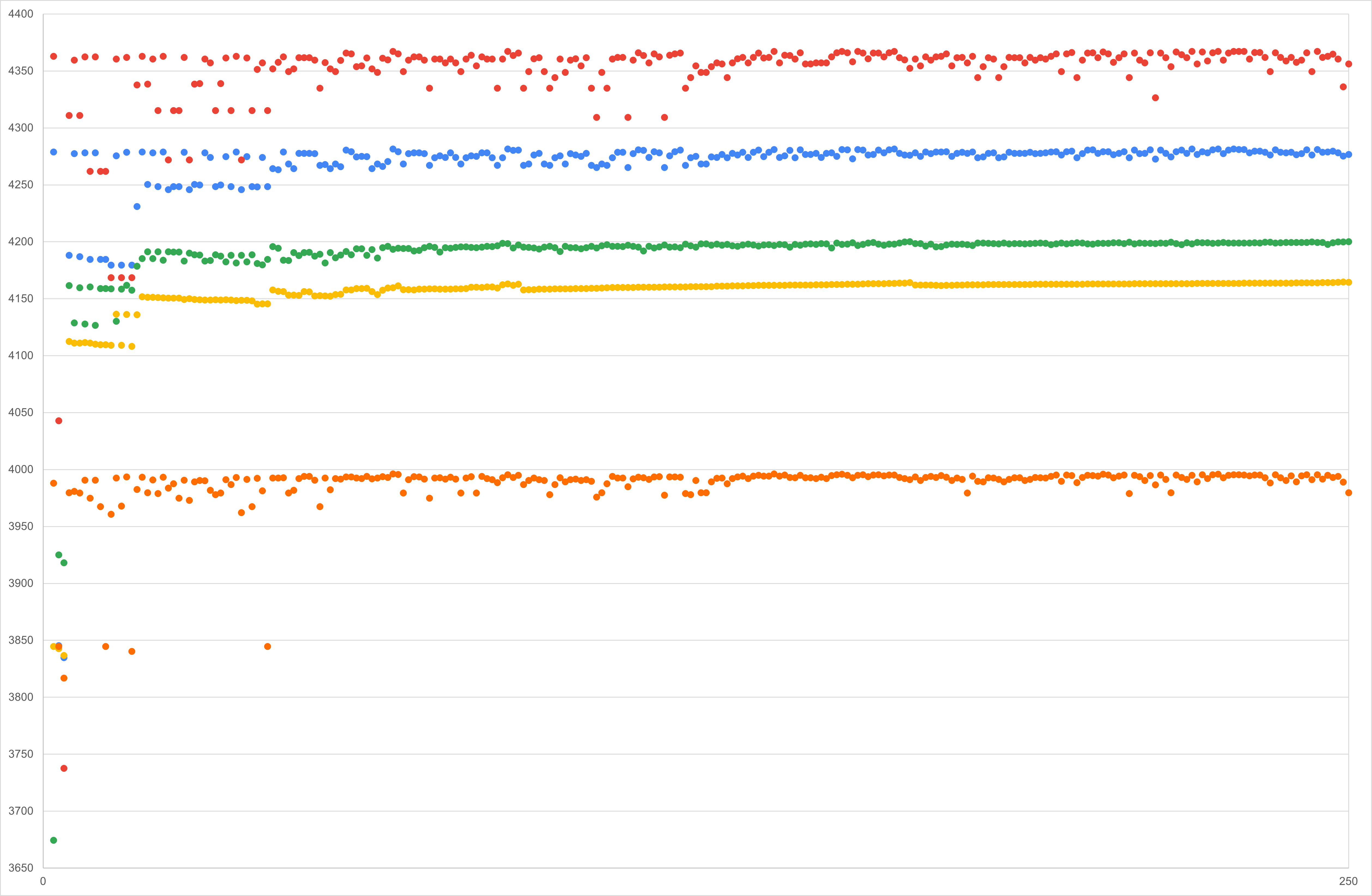}
\end{center}
\end{figure}

 In Table 1 of Section~\ref{sec:results}, we considered circular genomes with the instances $e=[1,2,3,4,5,6]$, $\sigma_1=[3,4,1,\overline{2},6,5]$, and $\sigma_2=[\overline{6},1,2,5,4,3]$, assuming dihedral symmetry. Alternatively, one could consider the genomes to have only $\cS_2$ symmetry (for example, now assume the genomes to have an origin of replication). Under this assumption, we further calculated the MLE distances between genomes with the above instances under each of the five models, and found an evolutionary signal in only two cases: an MLE of 5.20 between $\zee$ and $\zee\sigma_2$ under model (iii) and an MLE of 5.52 for the same genomes under model (iv). For reference, the values assuming circular symmetry are in Table~\ref{H6-app} below.

\begin{table}[h]
\centering
\caption{Pairwise distance estimates via MLE between genomes with instances $e=[1,2,3,4,5,6], \sigma_1=[3,4,1,\overline{2},6,5],\sigma_2=[\overline{6},3,4,5,2,1]$, assuming dihedral symmetry, under the five models.}\label{H6-app}
$
\begin{array}{c|c|c|c}
\textrm{model(s)} & d(\zee,\zee\sigma_1) & d(\zee,\zee\sigma_2) & d(\zee\sigma_1,\zee\sigma_2) \\
\hline 
 (i) & - & - & -\\
 (ii) & - & - & -\\
 (iii) & 8.659 & 8.81 & -\\
 (iv) & 6.94 & 4.254 & 7.01\\
 (v) & - & - & 8.097\\
\end{array}
$
\end{table}

\section{The 3-region case}\label{app:3reg}

To provide more detail on genomes, the genome algebras, the rearrangement models, and their construction under the different symmetry assumptions, we consider explicitly the case of genomes with three regions, where one can easily list all genomes. 

We again consider the genomes to have signed regions, so instances of genomes will be elements of the group $H_3$, which has $2^3 3! = 48$ elements. 
As above, we use one-line notation to represent these signed permutations. Recall that when representing genome instances as permutations, we are mapping regions to positions; specifically, representing a genome instance by $[a,b,\overline{c}]$ means that ``region 1 is in position $a$, region 2 is in position $b$ and region 3 is in position $c$ with negative orientation''. The same notation to represent a rearrangement or a symmetry is a mapping from {\em positions} to positions; specifically, applying $[a,b,\overline{c}]$ to a genome instance means that ``the region that was in position 1 moves to position $a$, the region that was in position 2 moves to position $b$ and the region that was in position 3 moves to position $c$ and reverses its orientation''.

Suppose firstly that the (circular) genomes have an origin of replication between positions $1$ and $3$, so the appropriate symmetry group is 
$Z = \{e,f\} = \{[1,2,3],[\overline{3},\overline{2},\overline{1}]\}$.
Thus there are $\tfrac{|H_3|}{|Z|} = 24$ distinct genomes here, with each genome having exactly two instances. For example, $[1,2,3]$ and $[\overline{3},\overline{2},\overline{1}]$ are both instances of the reference genome; $[1,3,\overline{2}]$ and $[\overline{3},\overline{1},2]$ are both instances of another genome. To represent these in the genome algebra, we set $\zee = \tfrac{1}{2}(e+f)$ and then the reference genome is $\zee = \tfrac{1}{2}\left([1,2,3]+[\overline{3},\overline{2},\overline{1}]\right)$; the second genome is 
$\zee  [1,3,\overline{2}] = \tfrac{1}{2}\left([1,3,\overline{2}]+ [\overline{3},\overline{1},2]\right)$.

We consider a model consisting of all inversions of one and two regions, equally weighted. Note that for three regions, this is {\em all} inversions. 
Formally, our model is
\[ \cM= \{ \zee [\overline{1},2,3]  \zee, \;\zee [1,\overline{2},3] \zee, \;\zee  [\overline{2},\overline{1},3] \zee \}\,,\]
with each rearrangement given a probability of $\tfrac{1}{3}$. Note that this covers all of the relevant rearrangements since $f\cdot[\overline{1},2,3]\cdot f = [1,2,\overline{3}]$; 
$\;f\cdot[\overline{2},\overline{1},3]\cdot f = [1,\overline{3},\overline{2}]$; and we disallow inversions over the origin of replication.

Then, performing calculations for distances of each genome from the reference, with the distance measures of minimum distance, MLE and MFPT, we obtained the results listed in Table~\ref{H3-lin}. (We note that for these dimensions, each set of calculations took a matter of seconds.) For each genome, we list a single instance as representative.

\begin{table}[hb]
\centering
\caption{Pairwise distance estimates between each genome with three oriented regions and the reference; genomes modelled with an origin of replication.}\label{H3-lin}
$\small{
\begin{array}{c|c|r|r}
\textrm{genome instance} & \textrm{min} & \mathrm{MLE} & \mathrm{MFPT} \\
\hline
{[1,\ 2,\ 3]} & 0 & 0.00 & 0.00 \\
{[1,\ 2,\ \overline{3}]} & 1 & 1.40 & 24.52 \\
{[1,\ 3,\ 2]} & 3 & 12.89 & 29.35 \\
{[1,\ 3,\ \overline{2}]} & 2 & 7.12 & 28.20 \\
{[1,\ \overline{2},\ 3]} & 1 & 1.20 & 19.50 \\
{[1,\ \overline{2},\ \overline{3}]} & 2 & 3.77 & 27.29 \\
{[1,\ \overline{3},\ 2]} & 2 & 7.12 & 28.20 \\
{[1,\ \overline{3},\ \overline{2}]} & 1 & 1.32 & 24.98 \\
{[2,\ 1,\ 3]} & 3 & 12.89 & 29.35 \\
{[2,\ 1,\ \overline{3}]} & 2 & - & 30.03 \\
{[2,\ 3,\ 1]} & 2 & - & 29.26 \\
{[2,\ 3,\ \overline{1}]} & 1 & 1.32 & 24.98 \\
{[2,\ \overline{1},\ 3]} & 2 & 7.12 & 28.20 \\
{[2,\ \overline{1},\ \overline{3}]} & 3 & - & 30.01 \\
{[2,\ \overline{3},\ 1]} & 3 & - & 30.01 \\
{[2,\ \overline{3},\ \overline{1}]} & 2 & 7.12 & 28.20 \\
{[3,\ 1,\ 2]} & 2 & - & 29.26 \\
{[3,\ 1,\ \overline{2}]} & 3 & - & 30.01 \\
{[3,\ 2,\ 1]} & 3 & - & 29.81 \\
{[3,\ 2,\ \overline{1}]} & 2 & 3.77 & 27.29 \\
{[3,\ \overline{1},\ 2]} & 3 & - & 30.01 \\
{[3,\ \overline{1},\ \overline{2}]} & 2 & - & 30.03 \\
{[3,\ \overline{2},\ 1]} & 2 & - & 29.11 \\
{[3,\ \overline{2},\ \overline{1}]} & 1 & 1.40 & 24.52
\end{array} }
$
\end{table}

The coincidences of values in the table are due to the genomes forming equivalence classes. There are, in this case, 12 distinct equivalence classes of genomes; members of any equivalence class are equidistant from the reference. For more details, we refer the reader to Section~\ref{sec:results} and the references therein. 

The regular representation of the model element in the genome algebra, that is, the Markov matrix representing rearrangement via this model, is

\[ \rho(\tilde{\ess}) = \scriptsize{\left(\begin{array}{rrrrrrrrrrrrrrrrrrrrrrrr}
0 & \frac{1}{6} & 0 & 0 & \frac{1}{3} & 0 & 0 & \frac{1}{6} & 0 & 0 & 0 & \frac{1}{6} & 0 & 0 & 0 & 0 & 0 & 0 & 0 & 0 & 0 & 0 & 0 & \frac{1}{6} \\
\frac{1}{6} & 0 & 0 & 0 & 0 & \frac{1}{3} & \frac{1}{6} & 0 & 0 & 0 & \frac{1}{6} & 0 & 0 & 0 & 0 & 0 & 0 & 0 & 0 & 0 & 0 & 0 & \frac{1}{6} & 0 \\
0 & 0 & 0 & \frac{1}{3} & 0 & \frac{1}{6} & \frac{1}{6} & 0 & 0 & 0 & 0 & 0 & \frac{1}{6} & 0 & 0 & 0 & 0 & 0 & 0 & 0 & 0 & \frac{1}{6} & 0 & 0 \\
0 & 0 & \frac{1}{3} & 0 & \frac{1}{6} & 0 & 0 & \frac{1}{6} & 0 & 0 & 0 & 0 & 0 & \frac{1}{6} & 0 & 0 & 0 & 0 & 0 & 0 & \frac{1}{6} & 0 & 0 & 0 \\
\frac{1}{3} & 0 & 0 & \frac{1}{6} & 0 & \frac{1}{6} & 0 & 0 & 0 & 0 & 0 & 0 & 0 & 0 & 0 & \frac{1}{6} & 0 & 0 & 0 & \frac{1}{6} & 0 & 0 & 0 & 0 \\
0 & \frac{1}{3} & \frac{1}{6} & 0 & \frac{1}{6} & 0 & 0 & 0 & 0 & 0 & 0 & 0 & 0 & 0 & \frac{1}{6} & 0 & 0 & 0 & \frac{1}{6} & 0 & 0 & 0 & 0 & 0 \\
0 & \frac{1}{6} & \frac{1}{6} & 0 & 0 & 0 & 0 & \frac{1}{3} & \frac{1}{6} & 0 & 0 & 0 & 0 & 0 & 0 & 0 & 0 & \frac{1}{6} & 0 & 0 & 0 & 0 & 0 & 0 \\
\frac{1}{6} & 0 & 0 & \frac{1}{6} & 0 & 0 & \frac{1}{3} & 0 & 0 & \frac{1}{6} & 0 & 0 & 0 & 0 & 0 & 0 & \frac{1}{6} & 0 & 0 & 0 & 0 & 0 & 0 & 0 \\
0 & 0 & 0 & 0 & 0 & 0 & \frac{1}{6} & 0 & 0 & \frac{1}{6} & 0 & 0 & \frac{1}{6} & 0 & 0 & \frac{1}{3} & 0 & 0 & 0 & \frac{1}{6} & 0 & 0 & 0 & 0 \\
0 & 0 & 0 & 0 & 0 & 0 & 0 & \frac{1}{6} & \frac{1}{6} & 0 & 0 & 0 & 0 & \frac{1}{6} & \frac{1}{3} & 0 & 0 & 0 & \frac{1}{6} & 0 & 0 & 0 & 0 & 0 \\
0 & \frac{1}{6} & 0 & 0 & 0 & 0 & 0 & 0 & 0 & 0 & 0 & \frac{1}{6} & 0 & \frac{1}{3} & \frac{1}{6} & 0 & 0 & 0 & 0 & 0 & \frac{1}{6} & 0 & 0 & 0 \\
\frac{1}{6} & 0 & 0 & 0 & 0 & 0 & 0 & 0 & 0 & 0 & \frac{1}{6} & 0 & \frac{1}{3} & 0 & 0 & \frac{1}{6} & 0 & 0 & 0 & 0 & 0 & \frac{1}{6} & 0 & 0 \\
0 & 0 & \frac{1}{6} & 0 & 0 & 0 & 0 & 0 & \frac{1}{6} & 0 & 0 & \frac{1}{3} & 0 & \frac{1}{6} & 0 & 0 & 0 & 0 & 0 & 0 & 0 & 0 & 0 & \frac{1}{6} \\
0 & 0 & 0 & \frac{1}{6} & 0 & 0 & 0 & 0 & 0 & \frac{1}{6} & \frac{1}{3} & 0 & \frac{1}{6} & 0 & 0 & 0 & 0 & 0 & 0 & 0 & 0 & 0 & \frac{1}{6} & 0 \\
0 & 0 & 0 & 0 & 0 & \frac{1}{6} & 0 & 0 & 0 & \frac{1}{3} & \frac{1}{6} & 0 & 0 & 0 & 0 & \frac{1}{6} & \frac{1}{6} & 0 & 0 & 0 & 0 & 0 & 0 & 0 \\
0 & 0 & 0 & 0 & \frac{1}{6} & 0 & 0 & 0 & \frac{1}{3} & 0 & 0 & \frac{1}{6} & 0 & 0 & \frac{1}{6} & 0 & 0 & \frac{1}{6} & 0 & 0 & 0 & 0 & 0 & 0 \\
0 & 0 & 0 & 0 & 0 & 0 & 0 & \frac{1}{6} & 0 & 0 & 0 & 0 & 0 & 0 & \frac{1}{6} & 0 & 0 & \frac{1}{3} & 0 & 0 & \frac{1}{6} & 0 & 0 & \frac{1}{6} \\
0 & 0 & 0 & 0 & 0 & 0 & \frac{1}{6} & 0 & 0 & 0 & 0 & 0 & 0 & 0 & 0 & \frac{1}{6} & \frac{1}{3} & 0 & 0 & 0 & 0 & \frac{1}{6} & \frac{1}{6} & 0 \\
0 & 0 & 0 & 0 & 0 & \frac{1}{6} & 0 & 0 & 0 & \frac{1}{6} & 0 & 0 & 0 & 0 & 0 & 0 & 0 & 0 & 0 & \frac{1}{6} & 0 & \frac{1}{6} & \frac{1}{3} & 0 \\
0 & 0 & 0 & 0 & \frac{1}{6} & 0 & 0 & 0 & \frac{1}{6} & 0 & 0 & 0 & 0 & 0 & 0 & 0 & 0 & 0 & \frac{1}{6} & 0 & \frac{1}{6} & 0 & 0 & \frac{1}{3} \\
0 & 0 & 0 & \frac{1}{6} & 0 & 0 & 0 & 0 & 0 & 0 & \frac{1}{6} & 0 & 0 & 0 & 0 & 0 & \frac{1}{6} & 0 & 0 & \frac{1}{6} & 0 & \frac{1}{3} & 0 & 0 \\
0 & 0 & \frac{1}{6} & 0 & 0 & 0 & 0 & 0 & 0 & 0 & 0 & \frac{1}{6} & 0 & 0 & 0 & 0 & 0 & \frac{1}{6} & \frac{1}{6} & 0 & \frac{1}{3} & 0 & 0 & 0 \\
0 & \frac{1}{6} & 0 & 0 & 0 & 0 & 0 & 0 & 0 & 0 & 0 & 0 & 0 & \frac{1}{6} & 0 & 0 & 0 & \frac{1}{6} & \frac{1}{3} & 0 & 0 & 0 & 0 & \frac{1}{6} \\
\frac{1}{6} & 0 & 0 & 0 & 0 & 0 & 0 & 0 & 0 & 0 & 0 & 0 & \frac{1}{6} & 0 & 0 & 0 & \frac{1}{6} & 0 & 0 & \frac{1}{3} & 0 & 0 & \frac{1}{6} & 0
\end{array}\right)}\,.\]

The corresponding graph, with edges of weight $\tfrac{1}{3}$ and $\tfrac{1}{6}$ coloured grey and black respectively is in Figure~\ref{graph-H3-lin} below.

\begin{figure}[ht]
\begin{center}
\caption{Rearrangement paths under an all inversions model for genomes with three oriented regions; genomes modelled with an origin of replication. Grey edges have weight $\tfrac{1}{3}$ and black $\tfrac{1}{6}$.}\label{graph-H3-lin}
\includegraphics[width=0.81\textwidth]{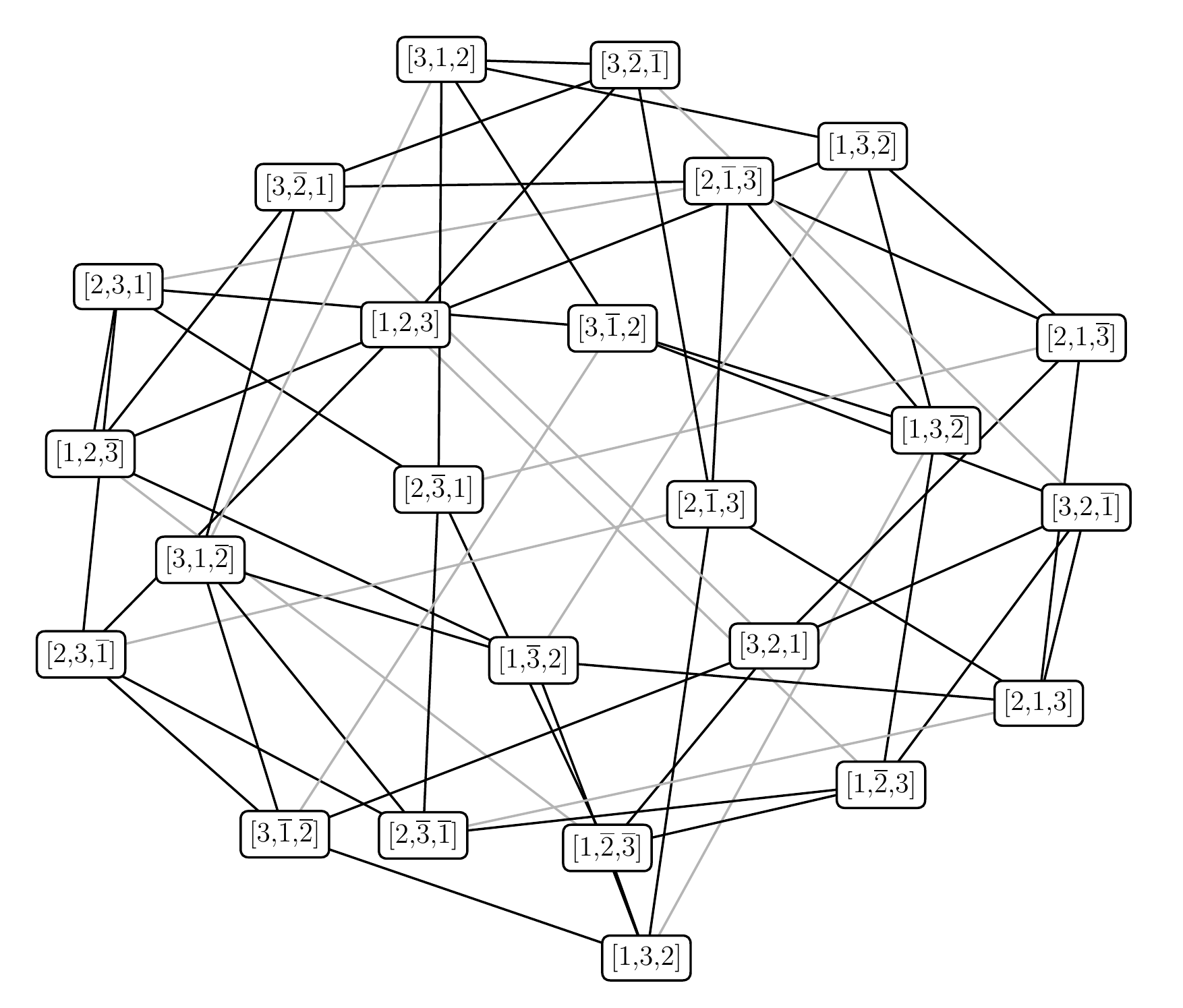}
\end{center}
\end{figure}

Alternatively, one can model the genomes without an origin of replication; we now turn to this case. The symmetry group now consists of all flips and rotations, that is,
\[ Z \cong \cD_3 = \{ [1,2,3], [2,3,1], [3,1,2],[\overline{3},\overline{2},\overline{1}], [\overline{1},\overline{3},\overline{2}],[\overline{2},\overline{1},\overline{3}]\}\,.\]
We define 
\[ \zee= \tfrac{1}{6} \sum_{z \in Z} z\,.\]
Accordingly, each genome has $6$ instances in $H_6$ and there are $\tfrac{|H_6|}{|Z|} = 8$ distinct genomes here. 

We wish to again consider a model consisting of all inversions of one and two regions, equally weighted. Now, one may obtain any one-region inversion from another by conjugation with a rotation (and, in general, this is true for any $k$-region inversion). Further, here the one- and two- region inversions are complementary: a two-region inversion is just a rotated flip of a one-region inversion. For example, whith the rotation $r=[2,3,1]$ and the flip, as above, $f=[\overline{3},\overline{2},\overline{1}]$, 
\[ r\cdot f\cdot [\overline{1},2,3] = [1,\overline{3},\overline{2}]\,.\]
Thus, the model is simply
\[  \cM= \{ \zee [\overline{1},2,3] \zee \}\,. \]

The distances of each genome from the reference, via the distance measures of minimum distance, MLE and MFPT, are listed in Table~\ref{H3-circ}. Again, there are coincidences in the values due to the equivalence classes of genomes; in this case, the genomes form four equivalence classes.

\begin{table}[h]
\centering
\caption{\small{Pairwise distance estimates between each genome with three signed regions and the reference; genomes modelled without an origin of replication.}}\label{H3-circ}
$
\begin{array}{c|c|r|r}
\textrm{genome instance} & \textrm{min} & \mathrm{MLE} & \mathrm{MFPT} \\
\hline
{[1,\ 2,\ 3]} & 0 & 0.00 & 0.0 \\
{[1,\ 2,\ \overline{3}]} & 1 & 1.65 & 7.0 \\
{[1,\ 3,\ 2]} & 3 &  -  & 10.0 \\
{[1,\ 3,\ \overline{2}]} & 2 & - & 9.0 \\
{[1,\ \overline{2},\ 3]}  & 1 & 1.65 & 7.0 \\
{[1,\ \overline{2},\ \overline{3}]}& 2 & - & 9.0 \\
{[1,\ \overline{3},\ 2]} & 2 & - & 9.0 \\
{[1,\ \overline{3},\ \overline{2}]}  & 1 & 1.65 & 7.0 \\
\end{array}
$
\end{table}

For this case, the regular representation of the model element in the genome algebra, that is, the Markov matrix representing rearrangement via this model, is

\[ \rho(\tilde{\ess}) =
\left(\begin{array}{rrrrrrrr}
0 & \frac{1}{3} & 0 & 0 & \frac{1}{3} & 0 & 0 & \frac{1}{3} \\
\frac{1}{3} & 0 & 0 & 0 & 0 & \frac{1}{3} & \frac{1}{3} & 0 \\
0 & 0 & 0 & \frac{1}{3} & 0 & \frac{1}{3} & \frac{1}{3} & 0 \\
0 & 0 & \frac{1}{3} & 0 & \frac{1}{3} & 0 & 0 & \frac{1}{3} \\
\frac{1}{3} & 0 & 0 & \frac{1}{3} & 0 & \frac{1}{3} & 0 & 0 \\
0 & \frac{1}{3} & \frac{1}{3} & 0 & \frac{1}{3} & 0 & 0 & 0 \\
0 & \frac{1}{3} & \frac{1}{3} & 0 & 0 & 0 & 0 & \frac{1}{3} \\
\frac{1}{3} & 0 & 0 & \frac{1}{3} & 0 & 0 & \frac{1}{3} & 0
\end{array}\right)\,.
\]

\begin{figure}[ht]
\begin{center}
\caption{Rearrangement paths under an all inversions model for genomes with three oriented regions; genomes modelled without an origin of replication.}\label{graph-H3-circ}
\includegraphics[width=0.54\textwidth]{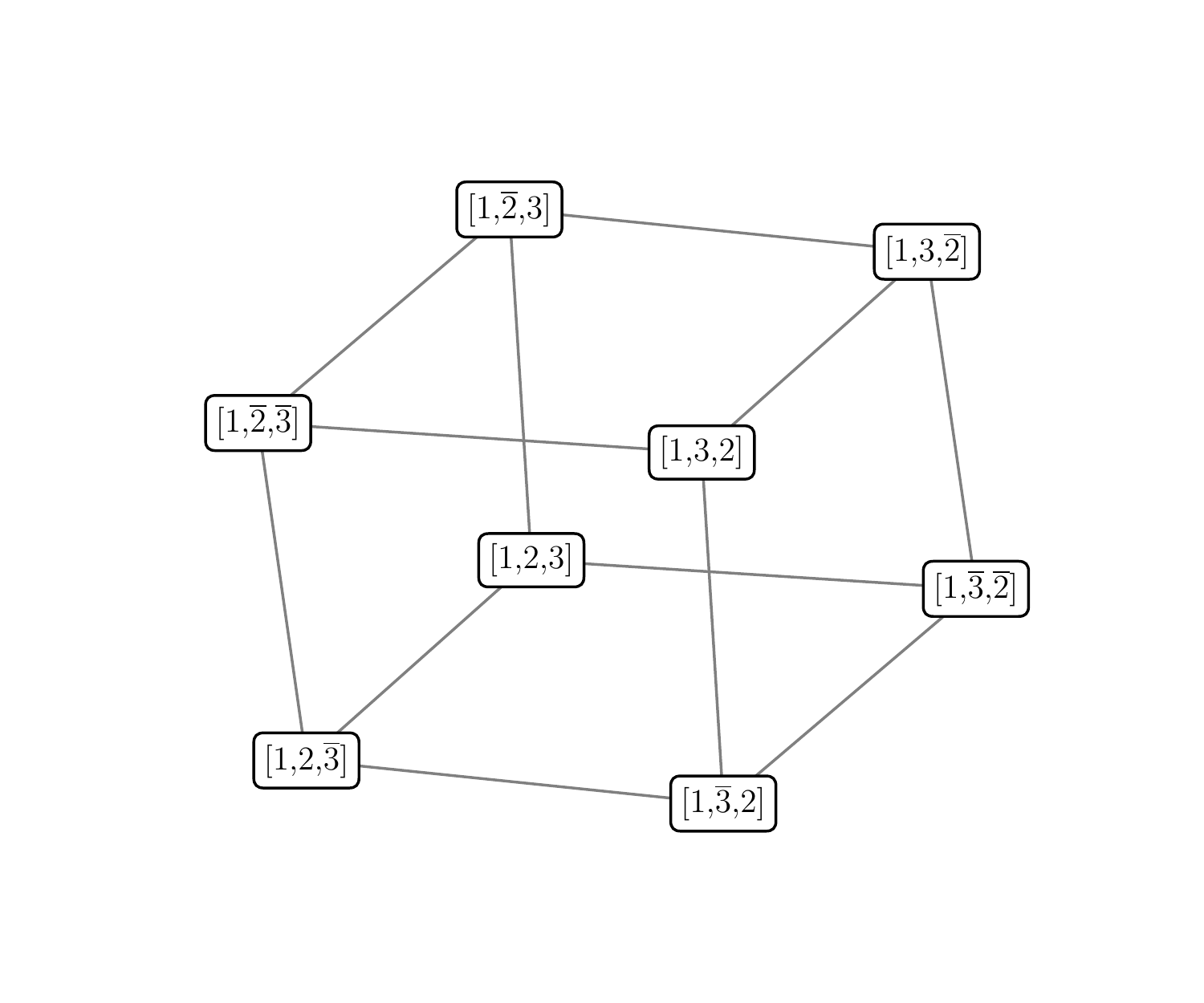}
\end{center}
\end{figure}

The corresponding graph is pictured in Figure~\ref{graph-H3-circ}.

\bibliographystyle{plain}
\bibliography{biblio}

\end{document}